# 基於橢圓曲線密碼學的同態雜湊演算法

Abel C. H. Chen

Chunghwa Telecom Co., Ltd.

chchen.scholar@gmail.com; ORCID 0000-0003-3628-3033

**摘要**

為解決資料明文在雲端運算環境曝露的問題，有部分的同態雜湊演算法被提出，可以計算每筆資料的雜湊值，並且僅在雲端運算環境儲存雜湊值，後續可對雜湊值進行計算和分析。然而，目前的同態雜湊演算法將可能隨著安全強度增加，將造成較長的產生雜湊值時間、總和雜湊值時間。因此，本研究提出基於橢圓曲線密碼學的同態雜湊演算法，通過橢圓曲線密碼學的特性建構同態雜湊的效果，並且以數學模型和實例來論證可行性。在實驗結果中，實作不同安全強度下，證明提出的演算法具備高效率。

*關鍵字：同態雜湊、橢圓曲線、密碼學。*

## 1. 前言

近年來，隨著雲端運算的蓬勃發展，許多企業已逐漸將其應用和服務部署到雲端運算環境。然而，在部署的同時，也需要考量到資訊安全等相關議題，避免企業資料或客戶資料在傳輸過程中被盜取或儲存在公雲環境時有資料洩露和曝露的風險。

有鑑於此，開始有同態加密(Homomorphic Encryption)演算法[1]-[4]、同態雜湊(Homomorphic Hashing)演算法[5]-[7]等被提出。這個系列的演算法可以允許終端設備對資料明文進行加密或產生雜湊值，再把密文或雜湊值儲存在雲端運算環境，避免資料明文在傳輸過程中或在雲端運算資料庫中曝露。其中，同態加密演算法可支持加法、乘法、以及搜尋等計算，可以在雲端運算環境對密文進行計算後，並把計算結果回傳至終端設備，由終端設備解密後得到明文進行加法、乘法、以及搜尋的結果。同態雜湊演算法可以讓雲端運算環境對雜湊值進行加總後的結果與明文加總後的雜湊值一致。

雖然目前已有部分的同態雜湊演算法被提出，但主要建構在指數運算的基礎上[5]-[7]，隨著指數值的增加和安全強度的增加，將造成較長的產生雜湊值時間、總和雜湊值時間。因此，本研究提出基於橢圓曲線密碼學(Elliptic Curve Cryptography, ECC)的同態雜湊演算法，通過橢圓曲線密碼學的特性建構同態雜湊的效果，並在離散對數問題(Discrete Logarithm Problem, DLP)的基礎上保障同態雜湊的安全性。

本論文共分為五個章節。第 2 節中介紹橢圓曲線密碼學，以及說明橢圓曲線密碼學的參數與基本計算方法。第 3 節提出本研究的基於橢圓曲線密碼學的同態雜湊演算法，並且運用數學模型證明原理、以及提供實例說明。第 4 節以不同的安全強度(Security Strength)等級下進行效能比較。第 5 節總結本研究貢獻和討論未來研究方向。

## 2. 橢圓曲線密碼學

橢圓曲線密碼學主要是建構在橢圓曲線函數上的非對稱式金鑰密碼學，以下從初始階段、產製金鑰階段、以及橢圓曲線密碼學特性分別說明。

### 2.1 初始階段

初始階段步驟包含：

(1). 定義橢圓曲線函數(如：NIST 標準定義的 Weierstrass 函數)，如公式(1)所示[8]。其中，將包含座標值(x, y)，以及係數 $a$ 和常數 $b$、質數模數 $p$。

$$y^2 = x^3 + ax^2 + b \mod p \qquad (1)$$

(2). 定義基點座標 $G$。其中，$G$ 點的座標值為$(G_x, G_y)$，例如：NIST 標準 P-224 中定義 $G$ 點的 $x$ 座標值 $G_x$ 為19277929113566293071110308034699480268319342194524401566497843520333、$G$ 點的 $y$ 座標值 $G_y$ 為19926808758034470970197974370888749184205991990603949537637343198772 [8]。

(3). 定義橢圓曲線函數的係數值、常數值、質數模數值。例如：NIST 標準 P-224 中定義係數 $a$ 值為26959946667150639794667015087019630673557916260026308143510066298878、常數 $b$ 值為18958286285566608000408668544493926415504680968679321075787234672564、質數模數 $p$ 值為26959946667150639794667015087019630673557916260026308143510066298881 [8]。

(4). 定義橢圓曲線點的加法計算方法。以 NIST 標準定義的 Weierstrass 函數為例，有兩個橢圓曲線上的點 $P_1 = (x_1, y_1)$和 $P_2 = (x_2, y_2)$，並且$P_1 \neq P_2$。則運用公式(2)、公式(3)、公式(4)可計算兩個橢圓曲線點相加後的座標值，$P_3$ (即$(x_3, y_3)$) = $P_1 + P_2$ [8]。

$$\lambda = \frac{y_2 - y_1}{x_2 - x_1} \quad (2)$$

$$x_3 = \lambda^2 - x_1 - x_2 \quad (3)$$

$$y_3 = \lambda(x_1 - x_3) - y_1 \quad (4)$$

(5). 定義橢圓曲線點的雙倍計算方法。以 NIST 標準定義的 Weierstrass 函數為例，有兩個橢圓曲線上的點 $P_1 = (x_1, y_1)$和 $P_2 = (x_2, y_2)$，並且$P_1 = P_2$ (即 $x_1 = x_2$ 且 $y_1 = y_2$)。則運用公式(5)、公式(6)、公式(7)可計算橢圓曲線點雙倍後的座標值，$P_3$ (即$(x_3, y_3)$) = $P_1 + P_2 = P_1 + P_1 = 2P_1$ [8]。

$$\lambda = \frac{3x_1^2 + a}{2y_1} \quad (5)$$

$$x_3 = \lambda^2 - 2x_1 \quad (6)$$

$$y_3 = \lambda(x_1 - x_3) - y_1 \quad (7)$$

### 2.2 產製金鑰階段

產製金鑰階段步驟包含：

(1). 選擇合適的橢圓曲線函數及其相關參數，例如：選擇 NIST 標準定義的 Weierstrass 函數和 P-224 等參數。

(2). 產製私鑰 $q$：隨機產生足夠大的整數作為私鑰 $q$。

(3). 產製公鑰 $Q$：計算 $Q = qG$，運用 $q$ 倍的基點座標 $G$ 點之 $x$ 座標值和作 $y$ 座標值作為公鑰。其中，採用公式(2)~公式(7)計算 $qG$。

### 2.3 橢圓曲線密碼學特性

本節整理橢圓曲線密碼學特性，條列如下：

(1). 當已知公鑰 $Q$，反推私鑰 $q$ 為離散對數問題，為 NP-Complete [8]。

(2). 當 $q_3 = q_1 + q_2$ 時，則 $q_3G = q_1G + q_2G$ (即 $Q_3 = Q_1 + Q_2$) [9]。

## 3. 研究方法

本研究主要提出基於橢圓曲線密碼學的同態雜湊演算法，提供同態雜湊應用。在 3.1 節中定義同態雜湊演算法的目標，在 3.2 節中描述本研究提出基於橢圓曲線密碼學的同態雜湊演算法，在 3.3 節中運用數學模型證明本研究提出的同態雜湊演算法，在 3.4 節中提供實例證明本研究提出的同態雜湊演算法。

### 3.1 同態雜湊演算法的目標

本節定義同態雜湊演算法的目標,假設有同態雜湊函數$f(\cdot)$,並符合公式(8)、公式(9)、公式(10)描述之特性。其中,第$i$筆資料明文為$p_i$、第$i$筆資料雜湊值為$h_{p_i}$、明文總和為$P$、雜湊值總和為$H$、資料數為$n$筆。並且,無法從雜湊值反推明文,以符合雜湊函數不可逆的特性。

$$h_{p_i} = f(p_i) \tag{8}$$

$$H = f(P), \text{where } P = \sum_{i=1}^{n} p_i \tag{9}$$

$$H = f(P) = \sum_{i=1}^{n} h_i = \sum_{i=1}^{n} f(p_i) \tag{10}$$

### 3.2 基於橢圓曲線密碼學的同態雜湊演算法

本節介紹本研究提出的基於橢圓曲線密碼學的同態雜湊演算法,主要結合橢圓曲線密碼學的特性產生雜湊值。其中,在操作上把第$i$筆資料明文為$p_i$作為橢圓曲線密碼學中的私鑰,再把對應的橢圓曲線密碼學中的公鑰作為第$i$筆資料雜湊值為$h_{p_i}$,如公式(11)所示。由於在橢圓曲線密碼學中,從公鑰(即雜湊值)反推私鑰(即明文)為離散對數問題,為NP-Complete,可以此視為不可逆特性。並且,根據橢圓曲線密碼學的特性,私鑰值(即明文)加總後對應的公鑰值(即雜湊值)(如公式(12)),與加總後的公鑰值(即雜湊值)相同(如公式(13))。因此,可證明本研究提出的基於橢圓曲線密碼學的同態雜湊演算法符合同態雜湊演算法的目標。

$$h_{p_i} = f(p_i) = p_i G \tag{11}$$

$$H = f(P) = f\left(\sum_{i=1}^{n} p_i\right) = \left(\sum_{i=1}^{n} p_i\right) G \tag{12}$$

$$H = \sum_{i=1}^{n} h_i = \sum_{i=1}^{n} p_i G \tag{13}$$

### 3.3 原理證明

由於公式(12)和公式(13)為符合同態雜湊演算法的目標的關鍵,本節以數學模型進行推導證明符合其特性。

以NIST標準定義的Weierstrass函數作為橢圓曲線函數為例,假設基點座標$G$為$(x_g, y_g)$,運用橢圓曲線點的雙倍計算方法(如公式(5)~公式(7)),可以取得$2G$座標為$(x_{2g}, y_{2g})$;其中,係數值和座標值結果如公式(14)~公式(16)所示。

$$\lambda_{2g} = \frac{3{x_g}^2 + a}{2 y_g} \tag{14}$$

$$x_{2g} = {\lambda_{2g}}^2 - 2x_g \tag{15}$$

$$y_{2g} = \lambda_{2g}(x_g - x_{2g}) - y_g \tag{16}$$

以下分別運用橢圓曲線點的加法計算方法(如公式(2)~公式(4))計算座標點$G_s$(即$(x_{3g,s}, y_{3g,s}) = G + 2G$)和另一個座標點$G_l$(即$(x_{3g,l}, y_{3g,l}) = 2G + G$),並證明此兩座標點為同一座標點(即$G_s = G_l = 3G$)。由此證明,當私鑰值(即明文)加總後對應的公鑰值(即雜湊值)(如公式(12)),與加總後的公鑰值(即雜湊值)相同(如公式(13))。

座標點$G_s$(即$(x_{3g,s}, y_{3g,s}) = G + 2G$)的係數值和座標值結果如公式(17)~公式(19)所示。

$$\lambda_{3g,s} = \frac{y_{2g} - y_g}{x_{2g} - x_g} \tag{17}$$

$$x_{3g,s} = {\lambda_{3g,s}}^2 - x_g - x_{2g} \tag{18}$$

$$y_{3g,s} = \lambda_{3g,s}(x_g - x_{3g,s}) - y_g \tag{19}$$

座標點$G_l$(即$(x_{3g,l}, y_{3g,l}) = 2G + G$)的係數值和座標值結果如公式(20)~公式(22)所示。

$$\lambda_{3g,l} = \frac{y_g - y_{2g}}{x_g - x_{2g}} \tag{20}$$

$$x_{3g,l} = {\lambda_{3g,l}}^2 - x_{2g} - x_g \tag{21}$$

$$y_{3g,l} = \lambda_{3g,l}(x_{2g} - x_{3g,l}) - y_{2g} \quad (22)$$

通過公式(23)可以證明係數值相同(即$\lambda_{3g,s} = \lambda_{3g,l}$)，並重新命名為$\lambda_{3g}$。

$$\lambda_{3g,s} = \frac{y_{2g} - y_g}{x_{2g} - x_g} = \frac{y_g - y_{2g}}{x_g - x_{2g}} = \lambda_{3g,l} = \lambda_{3g} \quad (23)$$

通過公式(24)和公式(25)可以證明座標點$G_s$和座標點$G_l$的 $x$ 座標值同為$\left(\frac{y_{2g}-y_g}{x_{2g}-x_g}\right)^2 - x_g - (\lambda_{2g}^2 - 2x_g)$，並重新命名為$x_{3g}$。

$$\begin{aligned} x_{3g,s} &= \lambda_{3g,s}^2 - x_g - x_{2g} \\ &= \lambda_{3g}^2 - x_g - x_{2g} \\ &= \left(\frac{y_{2g} - y_g}{x_{2g} - x_g}\right)^2 - x_g - (\lambda_{2g}^2 - 2x_g) \end{aligned} \quad (24)$$

$$\begin{aligned} x_{3g,l} &= \lambda_{3g,l}^2 - x_{2g} - x_g \\ &= \lambda_{3g}^2 - x_{2g} - x_g \\ &= \lambda_{3g}^2 - x_g - x_{2g} \\ &= \left(\frac{y_{2g} - y_g}{x_{2g} - x_g}\right)^2 - x_g - (\lambda_{2g}^2 - 2x_g) \end{aligned} \quad (25)$$

通過公式(26)和公式(27)可以證明座標點$G_s$和座標點 $G_l$ 的 $y$ 座標值同為$\left(\frac{y_g-y_{2g}}{x_g-x_{2g}}\right)(-x_{3g}) + \left(\frac{x_g y_{2g} - x_{2g} y_g}{x_{2g} - x_g}\right)$，並重新命名為$y_{3g}$。

$$\begin{aligned} y_{3g,s} &= \lambda_{3g,s}(x_g - x_{3g,s}) - y_g \\ &= \lambda_{3g}(x_g - x_{3g}) - y_g \\ &= \lambda_{3g}(-x_{3g}) + \lambda_{3g}(x_g) - y_g \\ &= \left(\frac{y_{2g} - y_g}{x_{2g} - x_g}\right)(-x_{3g}) \\ &\quad + \left(\frac{y_{2g} - y_g}{x_{2g} - x_g}\right)(x_g) - y_g \end{aligned} \quad (26)$$

$$\begin{aligned} &= \left(\frac{y_{2g} - y_g}{x_{2g} - x_g}\right)(-x_{3g}) \\ &\quad + \left(\frac{x_g y_{2g} - x_g y_g - x_{2g} y_g + x_g y_g}{x_{2g} - x_g}\right) \\ &= \left(\frac{y_{2g} - y_g}{x_{2g} - x_g}\right)(-x_{3g}) \\ &\quad + \left(\frac{x_g y_{2g} - x_{2g} y_g}{x_{2g} - x_g}\right) \end{aligned}$$

$$\begin{aligned} y_{3g,l} &= \lambda_{3g,l}(x_{2g} - x_{3g,l}) - y_{2g} \\ &= \lambda_{3g}(x_{2g} - x_{3g}) - y_{2g} \\ &= \lambda_{3g}(-x_{3g}) + \lambda_{3g}(x_{2g}) - y_{2g} \\ &= \left(\frac{y_g - y_{2g}}{x_g - x_{2g}}\right)(-x_{3g}) \\ &\quad + \left(\frac{y_g - y_{2g}}{x_g - x_{2g}}\right)(x_{2g}) \\ &\quad - y_{2g} \end{aligned}$$

$$\begin{aligned} &= \left(\frac{y_g - y_{2g}}{x_g - x_{2g}}\right)(-x_{3g}) \\ &\quad + \left(\frac{x_{2g} y_g - x_{2g} y_{2g} - x_g y_{2g} + x_{2g} y_{2g}}{x_g - x_{2g}}\right) \\ &= \left(\frac{y_g - y_{2g}}{x_g - x_{2g}}\right)(-x_{3g}) \\ &\quad + \left(\frac{x_{2g} y_g - x_g y_{2g}}{x_g - x_{2g}}\right) \\ &= \left(\frac{y_g - y_{2g}}{x_g - x_{2g}}\right)(-x_{3g}) \\ &\quad + \left(\frac{x_g y_{2g} - x_{2g} y_g}{x_{2g} - x_g}\right) \end{aligned} \quad (27)$$

因此，通過上述可推導$G_s = G_l = 3G$（即$(x_{3g,s}, y_{3g,s}) = (x_{3g,l}, y_{3g,l}) = (x_{3g}, y_{3g})$）。依此類推，可證當私鑰值(即明文)加總後對應的公鑰值(即雜湊值)(如公式(12))，與加總後的公鑰值(即雜湊值)相同(如公式(13))。

### 3.4 實例說明

本節以實例說明基於橢圓曲線密碼學的同態雜湊演算法，以 NIST 標準定義的 Weierstrass 函數

作為橢圓曲線函數為例，採用 P-224 中定義的橢圓曲線函數及其參數，以下以十六進位制表示方式進行說明。

假設有 3 筆資料明文分別為 0CDD5C、0A3E66、以及 0A8E20，分別如公式(28)、(29)、(30)所示。並且運用公式(11)進行雜湊後得到雜湊值，分別如公式(31)、(32)、(33)所示。其中，由於數字太長，以"\"進行換行表示。

$$p_1 = 0CDD5C \qquad (28)$$

$$p_2 = 0A3E66 \qquad (29)$$

$$p_3 = 0A8E20 \qquad (30)$$

$$\begin{aligned} h_1 &= f(p_1) \\ &= (3D52F972A9D70B38A3D6F58 \\ &\backslash 3DF55B885EB2959E818556250 \\ &\backslash 8007742A, 9A1FC185CB8598240 \\ &\backslash CF6856FBA844AECD2B288BEF9 \\ &\backslash 4B8BDDB5545597) \end{aligned} \qquad (31)$$

$$\begin{aligned} h_2 &= f(p_2) \\ &= (EDE39EAED72AB45A74A5A52 \\ &\backslash B460ADE8F7AF382196B470576 \\ &\backslash E7CF4180, 1E8E022828D1FAE98 \\ &\backslash 3BE6E7427E84E1613A53252CE \\ &\backslash 312375EA844A5F) \end{aligned} \qquad (32)$$

$$\begin{aligned} h_3 &= f(p_3) \\ &= (67670504C5E3592DFC82EF \\ &\backslash FCA4F145D8ECC32423151A07 \\ &\backslash 0A4ACB7158, 99C86BE114EA9 \\ &\backslash A7B77A37B17618651C865E79 \\ &\backslash 6AB2D786D73AD4EB4D1) \end{aligned} \qquad (33)$$

運用公式(13)將 3 筆雜湊值加總結果如公式(34)所示。並且，可以運用公式(12)明文加總後進行雜湊結果(如公式(35)所示)進行比較，結果顯示一致，證實基於橢圓曲線密碼學的同態雜湊演算法的可行性。

$$\begin{aligned} H &= \sum_{i=1}^{3} h_i = \sum_{i=1}^{3} f(p_i) \\ &= (1611C11FD083F3A3C92981B \\ &\backslash CE50B874D70B70A5CB17688F \\ &\backslash 648CBF486, 4ECEF092A95E67E \\ &\backslash 29D459E43F68169EE3A00478 \\ &\backslash DCF3C479EC1DF6EA4) \end{aligned} \qquad (34)$$

$$\begin{aligned} f(P) &= f\left(\sum_{i=1}^{3} p_i\right) = \left(\sum_{i=1}^{3} p_i\right) G = H \\ &= (1611C11FD083F3A3C92981B \\ &\backslash CE50B874D70B70A5CB17688F \\ &\backslash 648CBF486, 4ECEF092A95E67E \\ &\backslash 29D459E43F68169EE3A00478 \\ &\backslash DCF3C479EC1DF6EA4) \end{aligned} \qquad (35)$$

### 3.5 小結與討論

本文在 3.2 節提出基於橢圓曲線密碼學的同態雜湊演算法，並在 3.3 節從理論證明演算法可行性，以及在 3.4 中從實例證明演算法可行性。未來可將基於橢圓曲線密碼學的同態雜湊演算法部署於實際場域。在終端設備先對每筆資料執行本研究提出的基於橢圓曲線密碼學的同態雜湊演算法計算對應的雜湊值，並且將每筆資料的雜湊值儲存到雲端資料庫，避免明文曝露的風險。後續想計算多筆資料加總後的雜湊值時，可以通過雲端運算資源進行雜湊值的加總，並且雜湊值的加總值將與明文加總值的雜湊值一致。

此外，本節雖以 NIST 標準定義的 Weierstrass 函數作為橢圓曲線函數為例，採用 P-224 中定義的橢圓曲線函數及其參數。在實際部署時可根據安全強度選擇不同的橢圓曲線函數及其參數。本研究將在第 4 節驗證和分析不同安全強度下的執行效率。

## 4. 實證分析與討論

為實際驗證基於橢圓曲線密碼學的同態雜湊演算法之效率，本研究使用一台 Windows 10 企業版的電腦執行演算法，驗證產生雜湊值時間(如 4.1 節)、總和雜湊值時間(如 4.2 節)。其中，實驗使用的軟硬體詳細規格是 CPU Intel(R) Core(TM) i7-10510U、記憶體 8 GB、OpenJDK 18.0.2.1、以及函式庫 Bouncy Castle Release 1.72。

並且，為了比較在不同安全強度下的方法效率，本研究採用美國國家標準暨技術研究院所規範的安全強度標準，如表 1 所示[10]。其中，安全強度分共可為五個等級，並且取得每個等級對應橢圓曲線密碼學的參數來實作。

表 1 安全強度與橢圓曲線密碼學參數

| 安全強度 | 橢圓曲線密碼學參數 |
|---|---|
| 80 | 160 (採用 NIST P-192 實作) |
| 112 | 224 (採用 NIST P-224 實作) |
| 128 | 256 (採用 NIST P-256 實作) |
| 192 | 384 (採用 NIST P-384 實作) |
| 256 | 512 (採用 NIST P-521 實作) |

### 4.1 產生雜湊值時間

本節主要驗證不同安全強度下，採用本研究演算法所需的產生雜湊值時間，即計算公式(11)所需時間。其中，在實驗環境中將隨機產生 10,000 筆資料(即 $n = 10000$)，並對每一筆資料計算其雜湊值。表 2 為每筆資料進行雜湊計算的平均時間，時間單位為毫秒。由實驗結果顯示，在不同安全強度下，雜湊計算的平均時間約為 5 毫秒，雖然隨著安全強度增加，計算時間也略有增加，但並不顯著。因此，在實際部署時，建議可以採取較高的安全強度，如此也可以也較廣的雜湊值域。

### 4.2 總和雜湊值時間

本節主要驗證不同安全強度下，採用本研究演算法所需的總和雜湊值時間，即計算公式(13)所需時間。其中，在實驗環境中對已儲存在資料庫中的 10,000 筆雜湊值(即 $n = 10000$)進行加總，並產生加總後的雜湊值。表 3 為對 10,000 筆雜湊值進行總和的時間，時間單位為毫秒。由實驗結果顯示，在不同安全強度下，雜湊值總和的時間約為 64.7346 毫秒~85.2743 毫秒，也就是平均每加總一筆雜湊值所需時間僅為 0.00647346 毫秒~0.00852743 毫秒，將可符合實際上線系統對計算時間上的要求。即使在安全強度 256 時，加總一筆雜湊值所需時間僅為 0.00734115 毫秒。因此，同 4.1 節建議，在實際部署時，建議可以採取較高的安全強度。

表 2 產生雜湊值時間(單位：毫秒)

| 安全強度 | 產生雜湊值時間 |
|---|---|
| 80 | 5.67269117 |
| 112 | 5.71806574 |
| 128 | 5.71746361 |
| 192 | 5.84428082 |
| 256 | 5.87799408 |

表 3 總和雜湊值時間(單位：毫秒)

| 安全強度 | 總和雜湊值時間 |
|---|---|
| 80 | 65.5990 |
| 112 | 64.7346 |
| 128 | 72.0647 |
| 192 | 85.2743 |
| 256 | 73.4115 |

## 5. 結論與未來研究

本研究提出基於橢圓曲線密碼學的同態雜湊演算法，可以取得雜湊值加總後的值與明文加總後的雜湊值相同，提供同態雜湊應用。在 3.3 節提供數學模型證明其可行性，在 3.4 節提供實例說明本研究方法的可行性。並且通過實驗分析，在不同的安全強下，都能有較低的產生雜湊值時間、總和雜湊值時間。

## 參考文獻

# Homomorphic Hashing Based on Elliptic Curve Cryptography

Abel C. H. Chen
Chunghwa Telecom Co., Ltd.
chchen.scholar@gmail.com

**Abstract**

For avoiding the exposure of plaintexts in cloud environments, some homomorphic hashing algorithms have been proposed to generate the hash value of each plaintext, and cloud environments only store the hash values and calculate the hash values for future needs. However, longer hash value generation time and longer hash value summary time may be required by these homomorphic hashing algorithms with higher security strengths. Therefore, this study proposes a homomorphic hashing based on elliptic curve cryptography (ECC) to provide a homomorphic hashing function in accordance with the characteristics of ECC. Furthermore, mathematical models and practical cases have been given to prove the proposed method. In experiments, the results show that the proposed method have higher efficiency with different security strengths.

***Keywords***: *Homomorphic hashing, elliptic curve, cryptography*